\documentclass[aps,nofootinbib]{revtex4}
\usepackage[english]{babel}
\usepackage[utf8]{inputenc}
\usepackage[T1]{fontenc}
\usepackage{amsmath}
\usepackage{amsfonts}
\usepackage{amssymb}
\usepackage{amsthm}
\usepackage{graphicx}
\usepackage{array}
\usepackage{dcolumn}
\usepackage{epstopdf}
\usepackage{multirow}
\usepackage{floatflt}
\usepackage{underscore}
\usepackage{cellspace}
\usepackage{bm}
\usepackage{braket}
\usepackage{xcolor}

\usepackage{fancyhdr}

\begin{document}

\title{Diquark size effects in the quark-diquark approximation for baryons}

\author{Clara \surname{Tourbez}}
\email[E-mail: ]{clara.tourbez@umons.ac.be}
\thanks{ORCiD: 0009-0004-0909-6974}
\affiliation{Service de Physique Nucl\'{e}aire et Subnucl\'{e}aire,
Universit\'{e} de Mons,
UMONS Research Institute for Complex Systems,
Place du Parc 20, 7000 Mons, Belgium}

\author{Cyrille \surname{Chevalier}}
\email[E-mail: ]{cyrille.chevalier@umons.ac.be}
\thanks{ORCiD: 0000-0002-4509-4309}
\affiliation{Service de Physique Nucl\'{e}aire et Subnucl\'{e}aire,
Universit\'{e} de Mons,
UMONS Research Institute for Complex Systems,
Place du Parc 20, 7000 Mons, Belgium}

\author{Claude \surname{Semay}}
\email[E-mail: ]{claude.semay@umons.ac.be}
\thanks{ORCiD: 0000-0001-6841-9850}
\affiliation{Service de Physique Nucl\'{e}aire et Subnucl\'{e}aire,
Universit\'{e} de Mons,
UMONS Research Institute for Complex Systems,
Place du Parc 20, 7000 Mons, Belgium}

\begin{abstract}
Baryons can be described within several theoretical frameworks. Among them, the constituent approach is widely used. In this context, we aim to evaluate the accuracy of a particular model of baryons: the quark-diquark approximation. It consists in separating the three-body system into two subsequent two-body ones: a pair of two quarks, the diquark, and a second system consisting of the diquark and the third quark. This approximation is widely used, but its accuracy is rarely evaluated. The goal of this work is to perform this evaluation by comparing the quark-diquark model with a three-body model, both using the same semi-relativistic interaction. The baryon masses and some characteristic distances are computed and analysed within both approaches. Additionally, an original procedure to establish the quark-diquark potential will be presented with the aim to increase the precision of this approximation. It is shown that a diquark must not necessarily be compact to obtain good baryon masses.
\end{abstract}

\maketitle

\section{Introduction}

In the constituent quark model, baryons consist of three interacting quarks. If many baryon properties can be explained within such models, a lot of questions remain~\cite{klem10}. Among them is the possibility that two quarks form a compact cluster, a diquark, whose structure is relatively independent from the third quark. Since its introduction in 1964~\cite{gell64}, the quark-diquark picture has been widely used in hadronic physics, and it is still the key ingredient of recent works~\cite{carl08,gian09,sant15,maje16,gian19,guti19,torc23,farh23}. If this assumption seems physically relevant~\cite{anse93,bara21}, its limitations and range of validity are rarely fully investigated~\cite{flec88,sant19,zhu25}. Moreover, it is still debated whether a diquark should be considered as a fundamental object or simply as a convenient tool to compute baryon properties. It is within the framework of this latter approach that we place our work. To assess the relevance of the quark-diquark approximation to describe baryons, we compare this approach with a typical three-body model. The former consists in decomposing the three-body problem into two successive two-body subsystems. First, a group of two quarks forming the diquark is studied. Second, the interactions between this diquark and the remaining quark are analysed. In both approaches, the same QCD-inspired potential is used.

The goal of the present article is to evaluate the extend to which introducing a two-body substructure can accurately approximate a full three-body description. Baryons prove ideal candidates to test the diquark approximation for two main reasons. First, because these systems have been extensively studied; the literature contains abundant results for both three-body and quark-diquark potential models, offering an excellent basis for comparing the different practices. Second, the potentials used in baryon spectroscopy can be extended to describe other hadrons, such as multiquark or hybrid states. Thus, the conclusions drawn here regarding the accuracy of the quark–diquark approximation for baryons can likely be generalised to assess the relevance of using substructures in the study of the other hadrons. In summary, the aim of the present work is not to investigate baryons per se, but rather to use models from baryon spectroscopy as a fertile framework for evaluating the accuracy of the quark-diquark approximation. Concretely, the validity of this approximation is tested by relying on a three-body model that provides a good description of baryon spectra. More specifically, we are developing this study from several perspectives:
\begin{enumerate}
\item Explore three distinct approaches to describe the quark-diquark approximation, namely (i) a point-like diquark approximation \cite{flec88}, (ii) a convolution of the potential with the squared modulus of the diquark wave function \cite{gian09,gian19,carl08}, and (iii) a modified convolution involving a particle density operator.
\item Test its capacity to reproduce some characteristic lengths of baryons. More precisely, the distance between the two quarks composing the diquark and the distance between the diquark and the third quark are computed.
\item Evaluate its accuracy in predicting various baryon masses.
\end{enumerate} 
We only select baryons consisting of, at least, two up/down ($n$) or bottom ($b$) quarks to highlight the impact of constituent quark masses, without increasing the length of the paper. Although only light baryons with orbital angular momentum until $L=6$ are known \cite{klem10}, both ground and excited states with a total orbital angular momentum $L=8$ are considered to examine the impact of large orbital excitations and to compare with the results of \cite{flec88}. A quark-diquark configuration is expected to appear more naturally in asymmetric structures like $nnb$ or $bbn$ states, but it is nevertheless interesting to also look at baryons with three identical quarks, $nnn$ and $bbb$. A semi-relativistic dynamics, as in the point-form formalism, is adopted as it seems more physically relevant and well suited for baryon physics \cite{ferr11,sant15}.

The paper is organised as follows. The three-body model is first described in Sect.~\ref{sec: Three-body model}. The quark-diquark approximation is then explained in Sect.~\ref{sec: Quark-diquark approximation}. The potential describing the interactions between two quarks and between a quark and a diquark are respectively described in Sect.~\ref{subsec: Quark-quark interactions} and \ref{subsec: Quark-diquark interactions}. Sect.~\ref{sec: Results} starts with some explanations regarding the choice of quantum numbers in the quark-diquark approximation. Then, convolution formulas are compared in Sect.~\ref{subsec: Convolution} while the results regarding characteristic distances and masses are respectively given in Sect.~\ref{subsec: Characteristic distances} and \ref{subsec: Baryon masses}. Finally, some concluding remarks are given in Sect.~\ref{sec: Conclusion}. In the following, we work in natural units ($\hbar=c=1$).

\section{Three-body model} \label{sec: Three-body model}

We begin by solving Schrödinger-like equations in a three-body framework. Considering semi-relativistic dynamics, the baryon Hamiltonian, corresponding to the rest frame mass operator of the point-form formalism, is chosen to be
\begin{equation}
H=\sum_i \sqrt{\bm{p}_i^2+m_i^2}+ \sum_{i<j} V_{ij}(\bm{r}_{ij}),
\label{eq: Hamiltonian}
\end{equation}
where the indices $i,j \in \{1,2,3\}$ refer to the constituent quarks. The potential $V_{ij}$ describes the interactions between quarks $i$ and $j$, and depends on their relative position $\bm{r}_{ij}$. The explicit form of this potential is discussed in Sect.~\ref{subsec: Quark-quark interactions}.

To solve Schrödinger-like equations, we use an oscillator bases expansion which relies on the variational principle \cite{silv20, chev24}. Once the quark content is specified, along with the total angular momentum $L$, total spin $S$, total isospin $I$ and parity $P$, the baryon wave functions are obtained as linear combinations of harmonic oscillator eigenstates. The corresponding masses are upper bounds of the true eigenvalues, whose accuracy increases with the size of the basis. The functions within the oscillator bases are written in variables defined as
\begin{align}
\bm{\rho}&= \bm{r}_1-\bm{r}_2,\\
\bm{\lambda}&= \frac{m_1\bm{r}_1+m_2\bm{r}_2}{m_1+m_2}-\bm{r}_3.
\label{eq: rho and lambda}
\end{align}
The accuracy of the eigenvalues is also ruled by two non-linear variational parameters $a$ and $b$, defined as $\bm{\rho} = a \bm{x}$ and $\bm{\lambda} = b \bm{y}$, where $\bm{x}$ and $\bm{y}$ are dimensionless variables. Let us note that when particles $1$ and $2$ are identical, $\bm{\lambda}$ is the separation between the third quark and the centre of mass of the first two quarks in our semi-relativistic approach. This will always be the case in the following.

This choice of coordinates allows for the decomposition of the total baryonic excitation into a diquark-like excitation, including only the first two quarks, ruled by $\bm{\rho}$, and a quark-diquark-like excitation between this pair and the third quark, ruled by $\bm{\lambda}$. These coordinates lead to a very convenient notation for the states within the oscillator bases. One of these basis states is entirely characterised by a set of six quantum numbers:
\begin{equation}
(n_{\rho},l_{\rho},n_{\lambda},l_{\lambda},S_{12},I_{12} ),
\label{eq: 3body state notation}
\end{equation}
where $n_{\rho}$, $l_{\rho}$, $S_{12}$ and $I_{12}$ are respectively the radial, orbital, spin and isospin quantum numbers associated with the subsystem formed by the first two quarks in a $\bar{3}$ colour state, and $n_{\lambda}$ and $l_{\lambda}$ are the radial and orbital quantum numbers describing the excitation between this subsystem and the third quark. In Sect.~\ref{sec: Results}, this notation will allow a direct comparison with the quark-diquark approximation, assuming the diquark consists of the first two quarks.

\section{Quark-diquark approximation} \label{sec: Quark-diquark approximation}
The procedure applied to compute the baryon mass within the quark-diquark approximation is the usual one \cite{gian09,sant15}. It consists of two steps:
\begin{itemize}
\item We begin by solving the Schrödinger-like equations for the diquark in the centre of mass frame. The corresponding Hamiltonian is
\begin{equation}
H_D=\sqrt{\bm{p}^2+m_1^2}+\sqrt{\bm{p}^2+m_2^2}+V_{12},
\label{eq: HD}
\end{equation}
where $\bm p$ is the momentum conjugate to the relative distance between the first two constituent quarks and $V_{12}$ denotes the interaction potential between these quarks, discussed in the next section. This step provides the diquark mass $M_D$ and its internal wave function $\Psi_D$, as eigenvalue and eigenstate of $H_D$ respectively.
\item The second step consists of solving the Schrödinger-like equations for the quark-diquark system in the centre of mass frame. The associated Hamiltonian takes the form
\begin{equation}
H_B=\sqrt{\bm{p}^2+M_D^2}+\sqrt{\bm{p}^2+m_3^2}+V_{Dq},
\label{eq: HB}
\end{equation}
where $\bm p$ is the momentum conjugate to the relative distance between the diquark and the third quark and $V_{Dq}$ is the interaction potential between these two particles. The expression of $V_{Dq}$ requires some considerations, which are detailed in Sect.~\ref{subsec: Quark-diquark interactions}. This last step provides the baryon mass, eigenvalue of $H_B$
\end{itemize}

The Hamiltonians associated with the diquark and with the quark-diquark system will be solved thanks to the Lagrange mesh method. It is a powerful procedure to solve two-body problems\footnote{It has not been generalised to treat semi-relativistic three-body systems.}. Further information can be found in Sect.~\ref{subsubsec: Lagrange mesh application} and in \cite{baye15}.

\section{Interaction potentials} \label{sec: Interaction potentials}
Two distinct interaction potentials are used in the following to describe the baryons. The three-body model requires a potential describing the interactions between two quarks. This potential will also be used for the first step of the quark-diquark approximation, namely the study of the diquark. The second step of the quark-diquark approximation involves an interaction potential between the third quark and the diquark. Both potentials are introduced and discussed in detail in this section.

\subsection{Quark-quark interactions} \label{subsec: Quark-quark interactions}
The quark-quark interactions are described using a standard form. In particular, a potential inspired by the one introduced by Bhaduri, Cohler, and Nogami (BCN) \cite{bhad81} is employed. In this context, the quark-quark potential consists of two main components: a central-colour term of Cornell type and a spin-colour interaction term, whose short-range behaviour is represented by a Yukawa-type function. All colour factors are included in the parameters. Accordingly, the interaction potential between quarks $i$ and $j$ has the form
\begin{equation}
V_{ij}(r_{ij})=V_0+C r_{ij}-\frac{2b}{3r_{ij}}+\frac{\alpha_s}{9m_i m_j}\Lambda^2\; \frac{\mathrm{e}^{-\Lambda r_{ij}}}{r_{ij}} 4\; \bm{s}_i \cdot \bm{s}_j,
\label{eq: pot. q-q}
\end{equation}
where $\bm{s}_i$ is a spin operator. The only non-central part of this Hamiltonian is the spin-colour term. Although we will consider high orbital excitations for baryons, no colour-spin-orbit term is considered. This is a will to keep the potential as simple as possible, and because colour-spin-orbit interactions coming from the short- and long-range parts partly compensate \cite{luch91,caps86}. The parameters $V_0$, $C$, $b$, $\Lambda$ and $\alpha_s$ originate from \cite{theu01}, where the BCN potential is employed in combination with semi-relativistic kinematics. They were fitted to reproduce accurately the baryon spectra, and their values are listed in Table \ref{tab: parameters}. In addition to the original set of parameters, the bottom quark mass was fitted to the experimental mass of the $\Sigma_b$ baryon ($M_{\Sigma_b}=5.8135$ GeV), taken from \cite{meld08}. This fit was carried out within the three-body model using the oscillator bases expansion method described above.

{\renewcommand{\arraystretch}{2}
{\setlength{\tabcolsep}{5mm}
\begin{table}[!ht]
\centering
\begin{tabular}{cccccc}
\hline
$b=\alpha_s$ & $\Lambda$ (GeV) & $C$ ($\text{GeV}^2$) & $V_0$ (GeV) & $m_n$ (GeV) & $m_b$ (GeV) \\ \hline \hline
$0.57$       & $0.532$         & $0.121$              & $-0.409$    & $0.337$     & $5.415$    \\ \hline
\end{tabular}
\caption{Semi-relativistic parameters from the quark-quark interaction \eqref{eq: pot. q-q} from \cite{theu01}.}
\label{tab: parameters}
\end{table}}}

\subsection{Quark-diquark interactions} \label{subsec: Quark-diquark interactions}
Following the methodology given in Sect.~\ref{sec: Quark-diquark approximation}, the diquark masses are first computed, and the results are summarised in Table \ref{tab: diquark masses}. The spectroscopic notation is adopted as in \cite{gian09}. For instance, the ground state ($n=0$, $l=0$) is denoted by $1S$.

{\renewcommand{\arraystretch}{1.8}
{\setlength{\tabcolsep}{5mm}
\begin{table}[!ht]
\centering
\begin{tabular}{|c|c|c||c|c|c|}
\hline
Diquark & $l_D$ & Masse & Diquark & $l_D$ & Masse \\ \hline \hline
$\{bb\}_{1S}$ & 0  & $10.360$ & $[nn]_{1S}$ & 0   & $0.776$        \\ \hline
$\{bb\}_{2S}$ & 0  & $10.763$ & $\{nn\}_{1S}$ & 0 & $0.895$        \\ \hline
$\{bb\}_{3S}$ & 0  & $11.009$ & $[nn]_{2S}$ & 0   & $1.313$        \\ \hline
$\{bb\}_{4S}$ & 0  & $11.209$ & $\{nn\}_{2S}$ & 0 & $1.413$        \\ \hline
$[bb]_{1P}$ & 1    & $10.696$ & $[nn]_{1P}$ & 1   & $1.225$        \\ \hline
$[bb]_{1K}$ & 7    & $11.431$ & $\{nn\}_{1P}$ & 1 & $1.248$        \\ \hline
$\{bb\}_{1L}$ & 8  & $11.520$ & $\{nn\}_{1I}$ & 6 & $2.359$        \\ \hline
\end{tabular}
\caption{Various diquark masses (GeV). The notation $[..]$ ($\{..\}$) refers to a diquark spin $S_D=0$ ($S_D=1$). The subscripts indicate the excitations of diquarks using the standard spectroscopic notation \cite{gian09}, as specified by the angular momentum $l_D$ of the diquark.}
\label{tab: diquark masses}
\end{table}
}}

Various approaches exist to define the interaction potential between a quark and a diquark. In this subsection, three of them are presented, including an original one. The Lagrange mesh method, used in the quark-diquark approximation, is described with some detail on how it yields highly compact expressions for the computation of the quark-diquark potentials.

\subsubsection{Quark-diquark potential} \label{subsubsec: Quark-diquark potential}
Let us first discuss the colour properties of the diquarks. These subsystems must transform under the $\overline{3}$ representation in order to combine with a third quark into a colour singlet. In terms of colour, a diquark hence transforms in the same way as an antiquark. Moreover, it can be shown that, due to the colour factors, the quark-antiquark potential is connected to the quark-quark potential by the relation \cite{zhu25}
\begin{equation}
V_{q\overline{q}}=2 V_{qq},
\label{eq: V qq}
\end{equation}
where $V_{qq}$ is the quark-quark potential \eqref{eq: pot. q-q}.
A first approach consists in assuming that the quark-diquark potential has the same form as the quark-antiquark potential
\begin{equation}
V_{Dq}^{\text{unc}}=V_{q\overline{q}}=2 V_{qq}.
\label{eq: V unc}
\end{equation}
We refer to this potential as the unconvoluted one. Equations \eqref{eq: V qq} and \eqref{eq: V unc} are compatible with results from a holographic model for heavy baryons \cite{andr16}. The factor 2 multiplies the central-colour and the spin-colour parts because they are characterised by the same colour dependence. For the spin dependence of the spin-colour interaction, we choose a simple prescription which is a direct replacement of the spin of a quark with the spin of the diquark $\bm{S}_D$. As $\bm{S}_D = \bm{s}_1 + \bm{s}_2$, the spin-colour operator for the quark-diquark system is simply assumed to be $\bm{S}_D \cdot\bm{s}_3$. This simple choice is discussed in the conclusion. 
Let us note that, in \cite{flec88}, the diquark is assumed to be point-like with quarks $1$ and $2$ occupying the same place, without taking into account the global colour of the diquark. This led to a spin-spin interaction different from the one we obtain with our colour dependence.

This simple potential \eqref{eq: V unc} is then compared with two modified versions that incorporate the spatial extension of the diquark. The adopted procedure consists in convoluting the quark-antiquark potential with the diquark colour density. Let us first note that, as all quarks are in the same colour representation, the colour density can be identified as the quark density $\sigma$. In this work, two definitions of this density will be considered. The first employs the squared modulus of the diquark wave function. This convolution approach is commonly employed in diquark-like contexts \cite{gian09,gian19,carl08}, and is given by
\begin{equation}
V_{Dq1}(\bm{R})=\int |\psi_D(\bm{r})|^2\; V_{q\overline{q}}(|\bm{r}+\bm{R}|)\; \mathrm{d}^{3}r,
\label{eq: conv. int. 1}
\end{equation}
where $\bm{R}$ is the relative coordinate between the diquark and the third quark, as introduced in \cite{gian09}, and $\psi_D$ denotes the component of the diquark normalised wave function depending on the separation $\bm{r}$ between the two quarks. More precisely, the diquark wave function can be decomposed as
\begin{equation}
\Psi_D(\bm{r}_1, \bm{r}_2)=\psi_D(\bm{r})\phi_D(\bm{R}_{\text{CM}}),
\label{eq: diquark wave function}
\end{equation}
where $\phi_D$ depends on the centre of mass coordinates $\bm{R}_{\text{CM}}$.

\begin{figure}[!ht]
\centering
\includegraphics[scale=0.6]{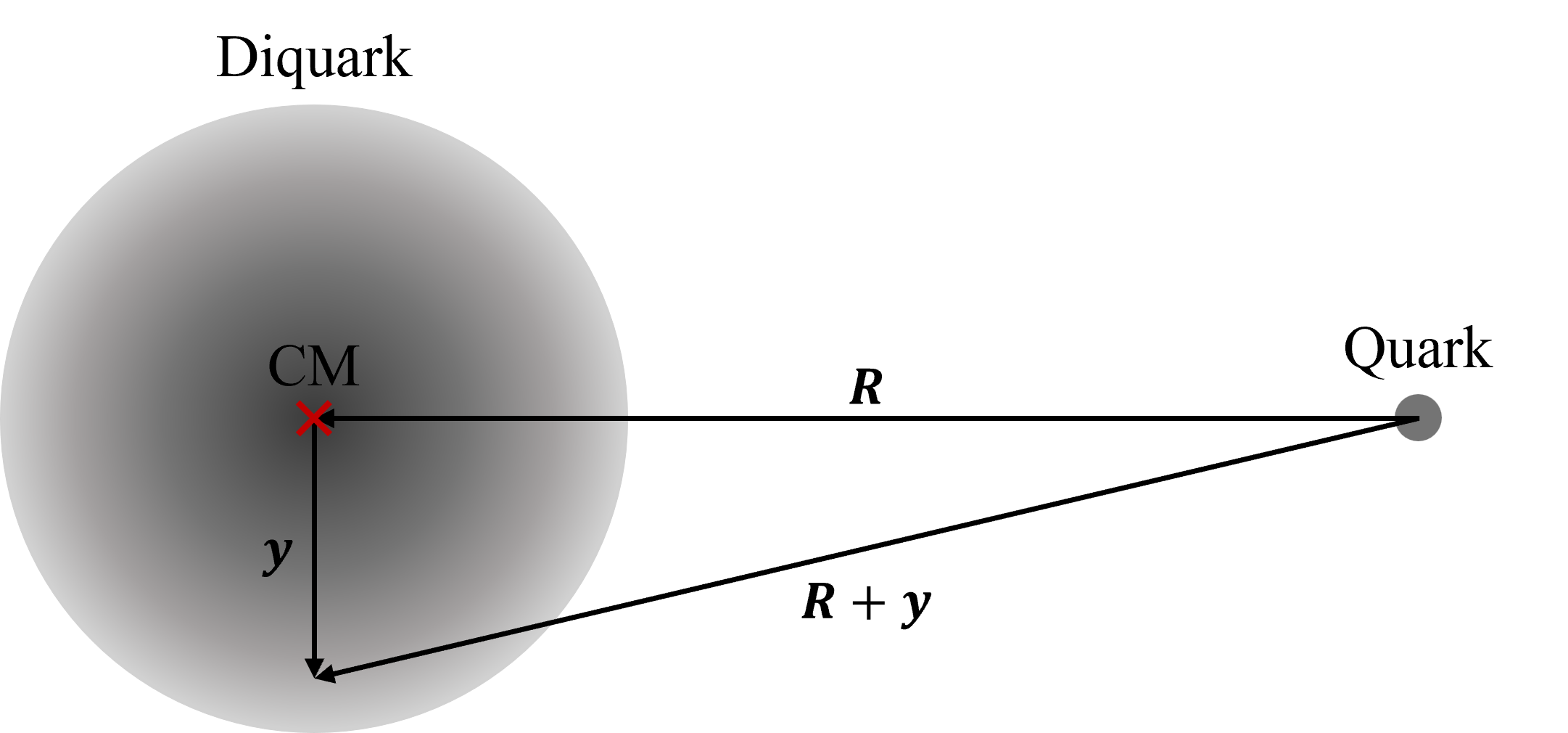}
\caption{Illustration of formula \eqref{eq: conv. int. 2}.}
\label{fig: convolution}
\end{figure}

The second expression includes a different definition of the quark density operator. The new convolution expression will therefore include this density, the expression of which is detailed below. The centre of mass frame is adopted for the following developments. In this context, the diquark wave function depends entirely on the relative position $\bm{x}_1$ between the two quarks
\begin{equation}
\bm{x}_1=\bm{y}_1-\bm{y}_2,
\label{eq: x1}
\end{equation}
$\bm{y}_i$ being the position of the $i$th quark relative to the diquark centre of mass. With the assumption that the diquark consists of two identical quarks, $\bm{y}_1$ and $\bm{y}_2$ become
\begin{align}
\bm{y}_1=\frac{\bm{x}_1}{2}, && \bm{y}_2=-\frac{\bm{x}_1}{2}.\label{eq: yi}
\end{align}
Now that these preliminary considerations have been set out, the second convoluted potential is assumed to be
\begin{equation}
V_{Dq2}(\bm{R})=\int \sigma(\bm{y})\; V_{q\overline{q}}(|\bm{y}+\bm{R}|)\; \mathrm{d}^{3}y,
\label{eq: conv. int. 2}
\end{equation}
where $\bm{R}$ is the position between the diquark centre of mass and the third quark, as illustrated in Fig. \ref{fig: convolution}. To determine the quark density $\sigma(\bm{y})$, the diquark wave function is first expressed as
\begin{equation}
\psi_D(\bm{x}_1)=\braket{\bm{x}_1 | \psi_D}.
\end{equation}
The quark density expression $\sigma(\bm{y})$ is then obtained by taking the mean value of the associated operator $\hat{\sigma}(\bm{y})$ on the diquark wave function.
This operator originates from \cite{naza13} and, in the diquark centre of mass frame, is of the form 
\begin{equation}
\hat{\sigma}(\bm{y})=\frac{1}{2}\left[ \int \ket{\bm{x}_1} \delta^3\left(\bm{y}-\frac{\bm{x}_1}{2}\right) \bra{\bm{x}_1}\; \mathrm{d}^3 x_1 + \int \ket{\bm{x}_1} \delta^3\left(\bm{y}+\frac{\bm{x}_1}{2}\right) \bra{\bm{x}_1}\; \mathrm{d}^3 x_1 \right],
\label{eq: density operator}
\end{equation}
where $\hat{\sigma}(\bm {y})$ is normalised to $1$, and the Dirac delta functions are used to localise each quark in the diquark. According to \eqref{eq: yi}, in the centre of mass frame, these two quarks are respectively located at $\bm{x}_1/2$ and $-\bm{x}_1/2$. The mean value of this operator is
\begin{equation}
\sigma(\bm{y})=\braket{\psi_D | \hat{\sigma}(\bm{y}) | \psi_D}= \frac{1}{2}\left[\int |\psi_D(\bm{x}_1)|^2 \delta^3\left(\bm{y}-\frac{\bm{x}_1}{2}\right)\; \mathrm{d}^3 x_1 + \int |\psi_D(\bm{x}_1)|^2 \delta^3\left(\bm{y}+\frac{\bm{x}_1}{2}\right)\; \mathrm{d}^3 x_1\right].
\label{eq: rho original}
\end{equation}
A change of variables is then performed, and after a few algebra, $\sigma(\bm{y})$ reduces to
\begin{equation}
\sigma(\bm{y})=8 |\psi_D(2 \bm{y})|^2.
\end{equation}
After substituting this expression in \eqref{eq: conv. int. 2}, the second convoluted potential finally becomes
\begin{equation}
V_{Dq2}(\bm{R})=8 \int |\psi_D(2\bm{y})|^2\; V_{q\overline{q}}(|\bm{y}+\bm{R}|)\; \mathrm{d}^{3}y.
\label{eq: conv. int. 2 final}
\end{equation}

\begin{figure}[!ht]
\centering
\includegraphics[scale=0.7]{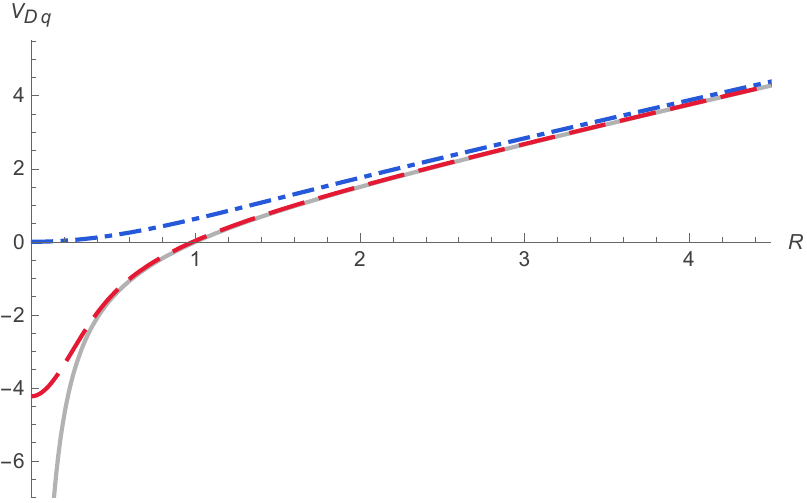}
\caption{Comparison of the three quark-diquark potentials for potential \eqref{eq: Vqqbar test} and wave function \eqref{eq: OH wf}. The solid grey curve represents the unconvoluted potential profile. The blue dash-dot curve corresponds to the first convoluted potential $V_{Dq1}$, while the red dashed curve illustrates the second convoluted potential $V_{Dq2}$. All quantities are given in arbitrary units.}
\label{fig: Comparison con and unconv}
\end{figure}

 To illustrate the behaviour of the three potentials \eqref{eq: V unc}, \eqref{eq: conv. int. 1} and \eqref{eq: conv. int. 2 final}, and the effect of the quark density on the quark-diquark potential, we consider a generic case with a simple quark-antiquark potential in arbitrary units, similar to the physical one presented at the beginning of this section
\begin{equation}
V_{q\overline{q}}(r)=-\frac{1}{r}+r.
\label{eq: Vqqbar test}
\end{equation}
Assuming that the diquark wave function is a ground-state harmonic oscillator ($n=0$ and $l=0$)
\begin{equation}
\Psi_{D}=\frac{\mathrm{e}^{-\frac{r^2}{2}}}{\pi^{\frac{3}{4}}},
\label{eq: OH wf}
\end{equation}
the resulting potentials are presented in Fig. \ref{fig: Comparison con and unconv}.
These three potentials differ mainly for small values of $R$. This observation is consistent with the fact that the further the third quark is from the diquark, the less the diquark spatial extension will have an impact on the interaction. The potential obtained with \eqref{eq: conv. int. 2 final} is closer to the unconvoluted one than the one obtained with \eqref{eq: conv. int. 1}. Potential \eqref{eq: conv. int. 2 final} being lower than potential \eqref{eq: conv. int. 1}, the corresponding mass values should be lower. The physical potentials presented at the beginning of this section show similar behaviours.

\subsubsection{Expressions of the convoluted potentials applied to the Lagrange mesh method} \label{subsubsec: Lagrange mesh application}
The expressions of the convoluted potentials can be simplified by considering the method of resolution, namely the Lagrange mesh method. Some of its characteristics are first introduced, as they will be useful in the following. 
The method relies on expanding the eigenstates $\ket{\Psi}$ in a basis of trial states $\ket{f_j}$,
\begin{equation}
\ket{\Psi}=\sum^N_{j=1} C_j \ket{f_j}.
\label{eq: trial state}
\end{equation}
The trial functions are chosen such that
\begin{equation}
\braket{\bm{r}|f_j}=\frac{f_j(r/h)}{\sqrt{h}r} Y_{lm}(\hat{r}),
\label{eq: trial fct}
\end{equation}
where $h$ is a scale parameter and $f_j$ denotes the $j$th regularised Lagrange function \cite{sema01}. 
This choice proves particularly convenient when associated with $N$-point Gauss-Laguerre quadratures \cite{abra64}
\begin{equation}
\int_0^{\infty} g(x)\mathrm{d}x \approx \sum^N_{k=1} \lambda_k g(x_k),
\label{eq: Gauss-Laguerre}
\end{equation}
where $x_k$ are mesh points defined as the roots of the $N$th Laguerre polynomial and $\lambda_k$ are the corresponding weights. One can show that the Lagrange functions satisfy the following property, named as the Lagrange condition,
\begin{equation}
f_j(x_k)=\lambda_k^{-1/2} \delta_{kj}.
\label{eq: lag. cond.}
\end{equation}
which greatly simplifies calculations within the Lagrange mesh method. This property will also be used to significantly simplify the formulas for the evaluation of the convoluted potentials, as described below.

As a first step to obtain these expressions, the diquark wave functions must be analysed. For diquarks with non-zero angular momentum ($L\neq0$), different angular momentum projections exist, and the wave function must be chosen accordingly. From all the possibilities, we choose to assume
\begin{equation}
\Psi_D(\bm{r})=\frac{1}{\sqrt{2l+1}} R_{nl}(r) \sum_{m=-l}^l Y_{lm}(\varphi,\theta),
\label{eq: diquark wf without m}
\end{equation}
where the radial wave function expression is obtained from equations \eqref{eq: trial state} and \eqref{eq: trial fct}:
\begin{equation}
R_{nl}(r)=\sum_j C_j \frac{f_j(r/h)}{\sqrt{h}r}.
\label{eq: radial wf}
\end{equation}
The assumption \eqref{eq: diquark wf without m} is motivated by the fact that any explicit dependence of the diquark wave function on the quantum number $m$ would also appear in the convoluted potentials. The third quark would then interact differently depending on the diquark orientation, given by $m$. Since no such phenomenon is observed in the baryon mass spectra, it seems reasonable to average over all possible $m$ values, giving the same statistical weight to all orientations. Furthermore, this assumption allows us to use a well-known property of the spherical harmonics \cite{vars88}:
\begin{equation}
\sum_{m=-l}^l |Y_{lm}(\varphi,\theta)|^2=\frac{2l+1}{4\pi}.
\label{eq: prop OH}
\end{equation}
This restores a kind of spherical symmetry for the diquark, which simplifies the convolution.
With these considerations, and by choosing the $z$ integration axis in the direction of $\bm{R}$, one obtains
\begin{equation}
|\bm{r}+\bm{R}|=\sqrt{R^2+r^2+2 R r \cos \theta}.
\end{equation}
The first convoluted potential then reads
\begin{equation}
V_{Dq1}(R)=\frac{1}{2}\int_0^{\infty} R_{nl}^2(r) I_V(r,R) r^2 \mathrm{d}r,
\label{eq: VDq1 int.}
\end{equation}
where
\begin{equation}
I_V(r,R)=\int_{-1}^{1} V_{q\overline{q}}\left(\sqrt{R^2+r^2+2R r \mu}\right)\; \mathrm{d}\mu.
\label{eq: Iv}
\end{equation}
The factor $1/2$ in Eq. \eqref{eq: VDq1 int.} results from the simplification of a $2\pi$ factor from the integral over $\varphi$ and a $4\pi$ factor from relation \eqref{eq: prop OH}. The factor $2l+1$ is removed by normalising the diquark wave function \eqref{eq: diquark wf without m}. Let us emphasise that a dependence on the angular momentum remains through the radial function $R_{nl}(r)$.

Applying an $N$-point Gauss-Laguerre quadrature \eqref{eq: Gauss-Laguerre}, the potential becomes
\begin{equation}
V_{Dq1}(R)\approx\frac{1}{2} \sum_{k=1}^N \lambda_k h^3 x_k^2 R_{nl}^2(h x_k) I_V(h x_k,R).
\end{equation}
The expression of the radial wave function \eqref{eq: radial wf} evaluated at a mesh point can be simplified using the Lagrange condition \eqref{eq: lag. cond.}. It takes the form
\begin{equation}
R_{nl}(h x_k)=\sum_j C_j \frac{f_j(x_k)}{\sqrt{h^3}x_k}=\frac{C_k}{h^{\frac{3}{2}}\lambda_k^{\frac{1}{2}}x_k}.
\end{equation}
Substituting this expression into the potential $V_{Dq1}$ yields
\begin{equation}
V_{Dq1}(R)\approx\frac{1}{2} \sum_{k=1}^N C_k^2 I_V(h x_k,R).
\label{eq: pot conv 1 lagmesh}
\end{equation}
A similar derivation applied to $V_{Dq2}(R)$ leads to
\begin{equation}
V_{Dq2}(R)\approx\frac{1}{2} \sum_{k=1}^N C_k^2 I_V\left( \frac{h x_k}{2},R \right).
\label{eq: pot conv 2 lagmesh}
\end{equation}
The evaluation of the integrals $I_V$ \eqref{eq: Iv} corresponding to the different components of the potential \eqref{eq: pot. q-q} yields analytical expressions, which are collected in Table \ref{tab: Iv}.
{\renewcommand{\arraystretch}{2.5}
{\setlength{\tabcolsep}{5mm}
\begin{table}[!ht]
\centering
\begin{tabular}{|c|cc|}
\hline
$V(r)$                               & \multicolumn{2}{c|}{$I_V(r,R)$}                                                 \\ \hline \hline
\multirow{2}{*}{$A r$}                        & \multicolumn{1}{c|}{$2A\frac{R^2+3r^2}{3r}$}                                 & $r\geq R$ \\ \cline{2-3} 
                                              & \multicolumn{1}{c|}{$2A\frac{r^2+3R^2}{3R}$}                                 & $r<R$     \\ \hline
\multirow{2}{*}{$-\frac{B}{r}$}               & \multicolumn{1}{c|}{$-\frac{2B}{r}$}                                         & $r\geq R$ \\ \cline{2-3} 
                                              & \multicolumn{1}{c|}{$-\frac{2B}{R}$}                                         & $r<R$     \\ \hline
\multirow{2}{*}{$D\, \frac{\mathrm{e}^{-Cr}}{r}$} & \multicolumn{1}{c|}{$D\, \frac{\mathrm{e}^{-C(r-R)}-\mathrm{e}^{-C(r+R)}}{CrR}$} & $r\geq R$ \\ \cline{2-3} 
                                              & \multicolumn{1}{c|}{$D\, \frac{\mathrm{e}^{C(r-R)}-\mathrm{e}^{-C(r+R)}}{CrR}$}  & $r<R$     \\ \hline
$F$                                           & \multicolumn{2}{c|}{$2F$}                                                           \\ \hline
\end{tabular}	
\caption{Analytical expressions of the integrals $I_V$ \eqref{eq: Iv} corresponding to the functions composing the interaction potential between two quarks \eqref{eq: pot. q-q}. $A$, $B$, $C$, $D$, and $F$ are constants.}
\label{tab: Iv}
\end{table}
}}
\section{Results} \label{sec: Results}

Several quark-diquark configurations are compatible for a fixed set of global quantum numbers $Q = \{L, S, I, P\}$. A possible procedure is to list by increasing mass values all the three-body and quark-diquark states compatible with $Q$ until a given energy. One can then identify each member of the two lists with the same order number. With this methodology, both spectra are obtained independently and only compared afterwards. However, we will see in Sec.~\ref{subsec: Baryon masses} that this choice does not necessarily result in an identification which relates states with similar internal structures. For that reason, and because part of our work consists in evaluating characteristic lengths, a different methodology is employed in the following.
 
The state notation introduced in Sect.~\ref{sec: Three-body model} with the oscillator bases expansion is also applicable to quark-diquark states, since the quantum numbers naturally split into diquark-like and quark-diquark-like excitations. So each quantum number set \eqref{eq: 3body state notation} for a three-body wave function can be associated with a quark-diquark state:
\begin{equation}
(n_D,l_D,n_{Dq},l_{Dq},S_D,I_D),
\label{eq: qD state notation}
\end{equation}
with
\begin{equation}
\begin{cases}
n_{\rho} \rightarrow n_D& \text{ radial excitation of the chosen diquark,}\\
l_{\rho} \rightarrow l_D& \text{ orbital excitation of the chosen diquark,}\\
n_{\lambda} \rightarrow n_{Dq}& \text{ radial excitation between the chosen diquark and the third quark,}\\
l_{\lambda} \rightarrow l_{Dq}& \text{ orbital excitation between the chosen diquark and the third quark,}\\
S_{12} \rightarrow S_D& \text{ spin of the chosen diquark,}\\
I_{12} \rightarrow I_D& \text{ isospin of the chosen diquark.}
\end{cases}
\label{Notation D12}
\end{equation}
Accordingly, the quantum numbers associated with the first two particles in the three-body model ($n_{\rho}$, $l_{\rho}$, $S_{12}$, and $I_{12}$) are reinterpreted as those of the diquark ($D$), while the quantum numbers of the (12)-3 system ($n_{\lambda}$ and $l_{\lambda}$) become those of the quark-diquark system ($Dq$).
This procedure ensures consistency between the two approaches and motivates the following procedure:
\begin{enumerate}
\item Solve the three-body equations for a baryon with fixed quark content, angular momentum $L$, spin $S$, isospin $I$, and parity $P$. This provides eigenstates expressed as linear combinations of basis states of the form \eqref{eq: 3body state notation}.
\item Identify the dominant states in the expansion, i.e., those with the largest coefficients.
\item Solve the quark-diquark equations for each of the selected dominant states. 
\end{enumerate}
With this identification methodology, the obtained quark-diquark states are compared to baryon states from the three-body calculation with a certain confidence about some similarities of their internal structure. At the end of Sec.~\ref{subsec: Baryon masses}, it is checked if this association is compatible with the identification based on the excitation level mentioned above.
Note that, in this procedure, the quark-diquark approximation does not consider a superposition of diquark states, whereas the three-body model solutions are a linear combination of different states for the two quarks associated with the diquark. Therefore, if the computed three-body eigenstate contains many components with significant coefficients, the comparison between the two approaches is not straightforward. 

\subsection{Convolution} \label{subsec: Convolution}

A comparison of the mass values obtained with the three-body and quark-diquark approaches is presented in Table \ref{tab: Comparison conv} for various types and states of baryons. The potentials $V_{Dq1}$ and $V_{Dq}^{\text{unc}}$ provide results with similar relative differences. However, in most cases, the masses predicted with $V_{Dq1}$ are larger than those obtained with the three-body model, whereas the masses computed with $V_{Dq}^{\text{unc}}$ tend to underestimate them. This latter behaviour was already reported in \cite{flec88}. Both of these observations are consistent with the behaviour of the potentials illustrated in Fig. \ref{fig: Comparison con and unconv}. The results show a significant improvement of the mass values when the convolution formula including the density operator \eqref{eq: density operator} is employed rather than the first formula. For this reason, we adopt the potential $V_{Dq2}$ to determine the baryon characteristics in the following sections. As discussed in Sect.~\ref{subsec: Quark-diquark interactions}, the use of $V_{Dq2}$, as given by \eqref{eq: conv. int. 2 final}, requires the assumption that the two quarks forming the diquark are identical. Since all baryons considered in this work contain at least two identical quarks, this condition is always satisfied. The observations arising from these results indicate that an appropriate convolution is necessary to reproduce the physical properties of the system with parameters coming from the three-body Hamiltonian.
{\renewcommand{\arraystretch}{2}
{\setlength{\tabcolsep}{3mm}
\begin{table}[!ht]
\centering
\begin{tabular}{|c|c|c|c|c|c|c|c|c|}
\hline
Baryon        & $I$                       & $L$ & $S$      & State  & $M\; \text{ 3-body}$ & $M\; (V_{Dq1})$ & $M (V_{Dq2})$ & $M\; (V_{Dq}^{\text{unc}})$ \\ \hline \hline
\multirow{3}{*}{$bbb$} & \multirow{3}{*}{$0$}           & $0$      & $\frac{3}{2}$ & $(0,0,0,0,1,0)^*$ & $14.410$                  & $14.898\; (3.4)$         & $14.570\; (1.1)$         & $13.823\; (4.1)$          \\ \cline{3-9} 
                       &                                & $8$      & $\frac{1}{2}$ & $(0,0,0,8,1,0)^*$ & \multirow{2}{*}{$16.488$} & $16.489\; (0.01)$        & $16.468\; (0.1)$        & $16.461\; (0.16)$         \\ \cline{3-5} \cline{7-9} 
                       &                                & $8$      & $\frac{1}{2}$ & $(0,7,0,1,0,0)$ &                           & $17.496\; (6.1)$         & $16.768\; (1.7)$        & $16.085\; (2.4)$          \\ \hline
\multirow{3}{*}{$bbn$} & \multirow{3}{*}{$\frac{1}{2}$} & $0$      & $\frac{3}{2}$ & $(0,0,0,0,1,0)$ & $10.224$                  & $10.398\; (1.7)$         & $10.249\; (0.2)$         & $10.063\; (1.6)$          \\ \cline{3-9} 
                       &                                & $0$      & $\frac{1}{2}$ & $(0,0,0,0,1,0)^*$ & $10.188$                  & $10.378\; (1.9)$         & $10.217\; (0.3)$         & $9.978\; (2.1)$           \\ \cline{3-9} 
                       &                                & $8$      & $\frac{1}{2}$ & $(0,8,0,0,1,0)^*$ & $11.922$                  & $12.776\; (7.2)$         & $12.041\; (1.0)$         & $11.127\; (6.7)$          \\ \hline
\multirow{4}{*}{$nnb$} & \multirow{2}{*}{$1$}           & $0$      & $\frac{3}{2}$ & $(0,0,0,0,1,1)$ & $5.837$                   & $6.483\; (11.1)$          & $6.036\; (3.4)$          & $5.497\; (5.8)$           \\ \cline{3-9} 
                       &                                & $0$      & $\frac{1}{2}$ & $(0,0,0,0,1,1)$ & $5.812$                   & $6.480\; (11.5)$          & $6.027\; (3.7)$          & $5.442\; (6.4)$           \\ \cline{2-9} 
                       & \multirow{2}{*}{$0$}           & $0$      & $\frac{1}{2}$ & $(0,0,0,0,0,0)^*$ & $5.583$                   & $6.215\; (11.3)$          & $5.850\; (4.8)$          & $5.406\; (3.2)$           \\ \cline{3-9} 
                       &                                & $8$      & $\frac{1}{2}$ & $(0,0,0,8,0,0)^*$ & $7.638$                   & $7.993\; (4.6)$          & $7.892\; (3.3)$          & $7.857\; (2.9)$           \\ \hline
\multirow{5}{*}{$nnn$} & $\frac{3}{2}$                  & $0$      & $\frac{3}{2}$ & $(0,0,0,0,1,1)$ & $1.226$                   & $1.661\; (35.5)$         & $1.302\; (6.2)$          & $1.061\; (13.5)$          \\ \cline{2-9} 
                       & \multirow{4}{*}{$\frac{1}{2}$} & $0$      & $\frac{1}{2}$ & $(0,0,0,0,0,0)^*$ & \multirow{2}{*}{$0.954$}  & $1.397\; (46.4)$         & $1.100\; (15.3)$         & $0.866\; (9.2)$           \\ \cline{3-5} \cline{7-9} 
                       &                                & $0$      & $\frac{1}{2}$ & $(0,0,0,0,1,1)$ &                           & $1.614\; (69.2)$         & $1.192\; (24.9)$         & $0.488\; (48.8)$          \\ \cline{3-9} 
                       &                                & $8$      & $\frac{1}{2}$ & $(0,0,0,8,0,0)^*$ & \multirow{2}{*}{$3.479$}  & $3.723\; (7.0)$          & $3.642\; (4.7)$          & $3.615\; (3.9)$           \\ \cline{3-5} \cline{7-9} 
                       &                                & $8$      & $\frac{1}{2}$ & $(0,6,0,2,1,1)$ &                           & $4.963\; (42.7)$         & $3.897\; (12.0)$          & $3.304\; (5.0)$           \\ \hline
\end{tabular}
\caption{Mass comparison (in GeV) of different states for the baryons $bbb$, $bbn$, $nnb$ and $nnn$ computed with the three-body and quark-diquark approaches. Columns 2 to 4 respectively show the baryon total isospin, angular momentum and spin. The states in the fifth column are those having the highest coefficient in the three-body expansion (see Sect.~\ref{sec: Results}). The sixth column indicates the masses computed with the three-body model, the seventh those computed with the quark-diquark approximation and $V_{Dq1}$, the eighth those computed with $V_{Dq2}$, and the last one with $V_{Dq}^{\text{unc}}$. The relative differences $(\%)$ between the two models are indicated in parentheses. For each ground state with $L=0$ and $L=8$, the state whose mass computed with $V_{Dq2}$ is closest to that obtained with the three-body model is marked with an asterisk ($^*$). These states are used in Table \ref{tab: distance errel}.}
\label{tab: Comparison conv}
\end{table}
}}

\subsection{Characteristic distances} \label{subsec: Characteristic distances}
The internal structure of baryons computed within a three-body model is carefully studied in \cite{flec88} with the aim of detecting diquark clustering. In particular, characteristic mean distances between quarks are computed but no comparison is made with the corresponding distances in the quark-diquark approach. So, to gain more insight on the validity of the quark-diquark model for describing the internal structure of the studied baryons, two characteristic distances are computed and compared in both approaches. Taking into account the definitions of the variables in the three-body and quark-diquark systems, it seems relevant to compare the mean distance between the first two quarks within the three-body system, $\langle \rho\rangle$, with the size of the diquark, $r_{qq}$, defined as the mean value of the radial variable in \eqref{eq: HD}. Similarly, the mean distance between the centre of mass of the subsystem formed by the first two quarks and the third quark within the three-body system, $\langle \lambda\rangle$, can be compared with the distance between the quark and the diquark, $r_{Dq}$, defined as the mean value of the radial variable in \eqref{eq: HB}. These computations are carried out for the states marked with an asterisk in Table \ref{tab: Comparison conv}, and the relative differences between both approaches are summarised in Table \ref{tab: distance errel}. The complete set of quantitative results is provided in Appendix \ref{app: Characteristic distances results}. The results clearly show that the quark-diquark approximation cannot naively be used for the computation of these types of observable mean values. This does not imply, however, that more physical observables, such as decay channel widths, cannot be accurately reproduced. Such calculations would require a redefinition of the operators and the wave functions to properly account for the peculiarities of the quark-diquark structure, in particular the size of the diquark. This could be done in the point-form formalism \cite{ferr11,des11,sant15}.

{\renewcommand{\arraystretch}{1.8}
{\setlength{\tabcolsep}{5mm}
\begin{table}[!ht]
\centering
\begin{tabular}{|c|c|c|c|}
\hline
System                 & Observable    & Ground state & $L=8$ \\ \hline \hline
\multirow{2}{*}{$bbb$} & $\Delta_{qq}$ & $-$0.42         & 0.68 \\
                       & $\Delta_{Dq}$ & $-$0.19         & $-$0.47  \\ \hline
\multirow{2}{*}{$bbn$} & $\Delta_{qq}$ & $-$0.16         & $-$0.26  \\
                       & $\Delta_{Dq}$ & $-$0.06         & $-$0.06  \\ \hline
\multirow{2}{*}{$nnb$} & $\Delta_{qq}$ & $-$0.64         & $-$0.40  \\
                       & $\Delta_{Dq}$ & $-$0.39         & $-$0.09  \\ \hline
\multirow{2}{*}{$nnn$} & $\Delta_{qq}$ & $-$0.28         & 0.53  \\
                       & $\Delta_{Dq}$ & $-$0.26         & $-$0.53 \\ \hline
\end{tabular}
\caption{Relative differences $\Delta_{qq}=(\braket{\rho}-r_{qq})/\braket{\rho}$ and $\Delta_{Dq}=(\braket{\lambda}-r_{Dq})/\braket{\lambda}$ between the characteristic distances obtained with the three-body and quark-diquark approaches with $V_{Dq2}$. The chosen states are those marked with an asterisk in Table \ref{tab: Comparison conv}. All the computed results can be found in Appendix \ref{app: Characteristic distances results}.}
\label{tab: distance errel}
\end{table}
}}

\begin{figure}[!ht]
\centering
\includegraphics[scale=0.5]{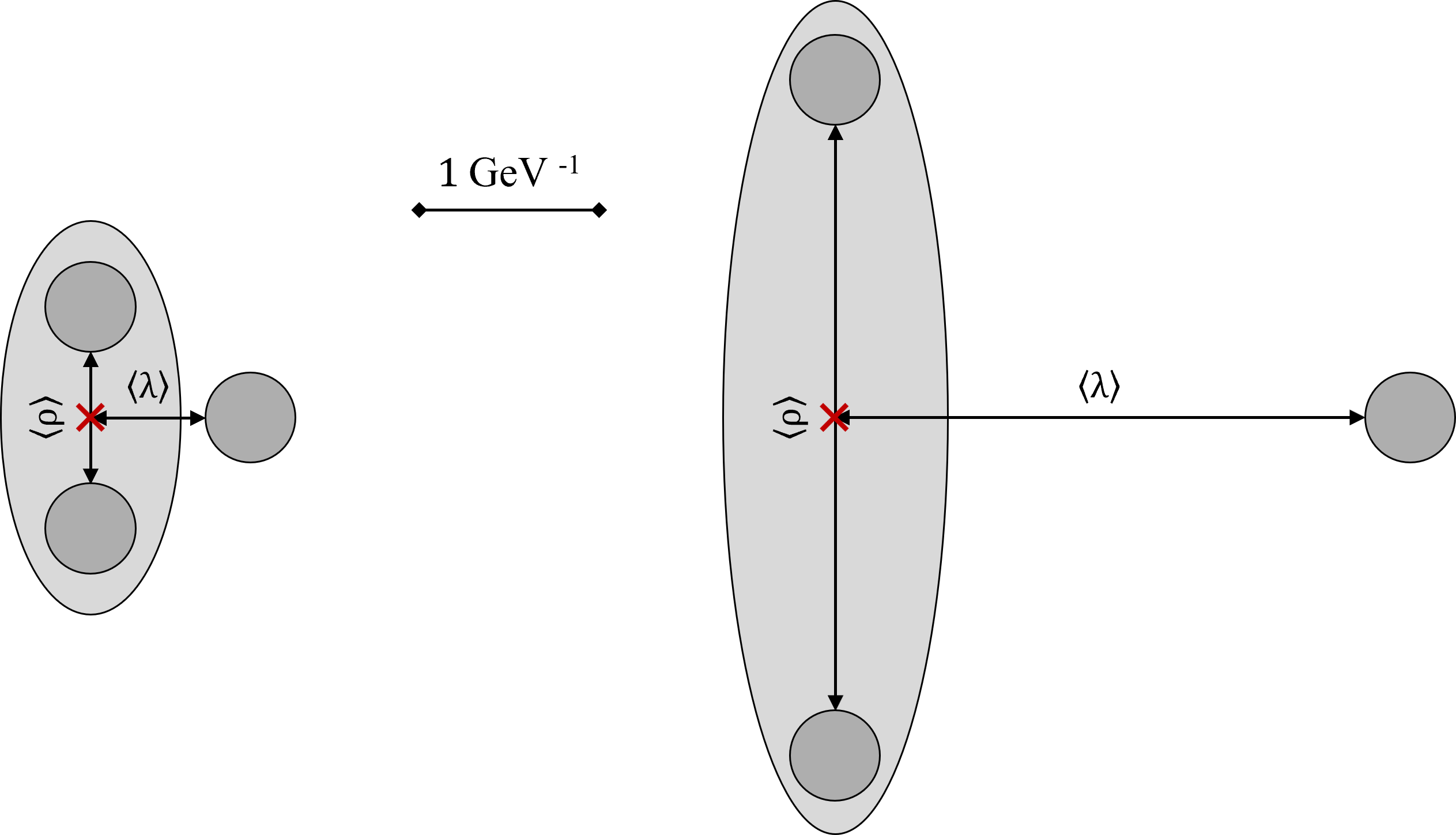}
\caption{Structures of the ground (left) and $L=8$ (right) states of the $bbb$ baryon obtained by computing the mean values of $\rho$ and $\lambda$. Both chosen states have a total spin $S=3/2$. The ratio defined in \eqref{eq: eta} takes the values $\eta=1.152$ and $\eta=1.105$ for these states, respectively.}
\label{fig: bbb fund&L=8}
\end{figure}

\begin{figure}[!ht]
\centering
\includegraphics[scale=0.5]{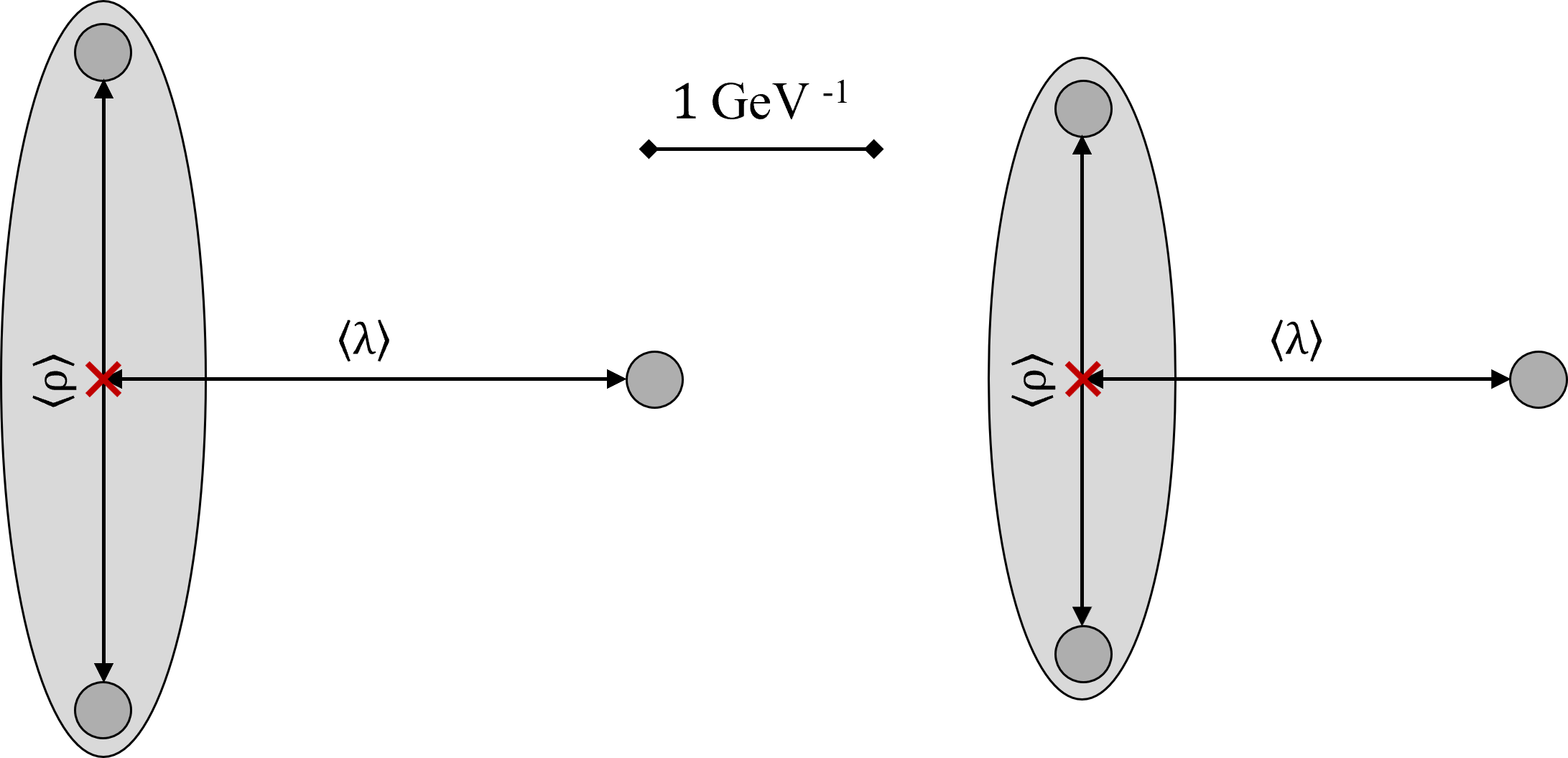}
\caption{Structures of the ground $S=3/2$ (left) and $S=1/2$ (right) states of the $nnn$ baryon obtained by computing the mean values of $\rho$ and $\lambda$. The ratio defined in \eqref{eq: eta} takes the values $\eta=1.153$ and $\eta=1.151$ for these states, respectively.}
\label{fig: nnn fund}
\end{figure}

\begin{figure}[!ht]
\centering
\includegraphics[scale=0.5]{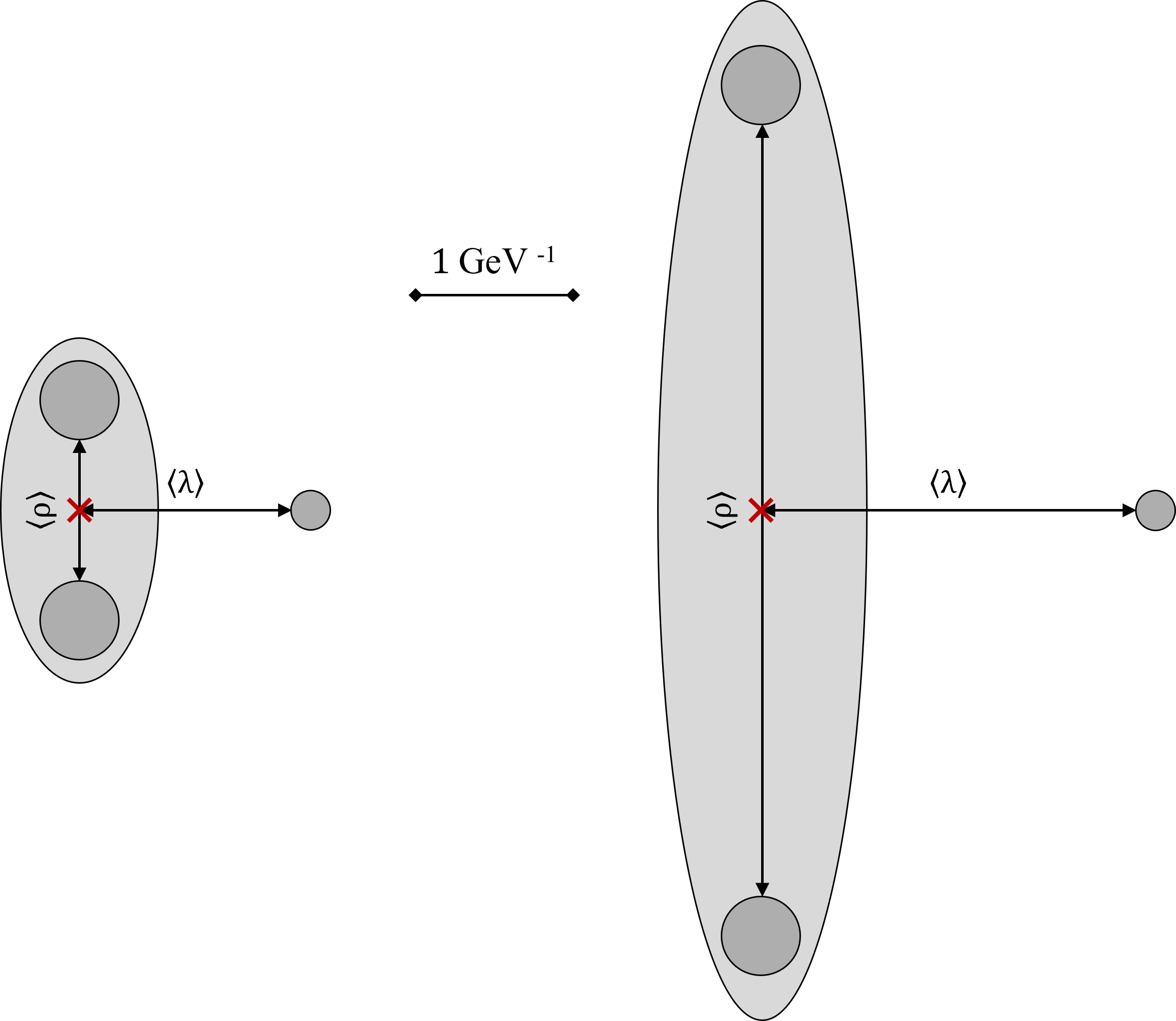}
\caption{Structures of the ground (left) and $L=8$ (right) states of the $bbn$ baryon obtained by computing the mean values of $\rho$ and $\lambda$. Both chosen states have a total spin $S=1/2$, and a total isospin $I=1/2$. The ratio defined in \eqref{eq: eta} takes the values $\eta=0.669$ and $\eta=2.056$ for these states, respectively.}
\label{fig: bbn fund&L=8}
\end{figure}

\begin{figure}[!ht]
\centering
\includegraphics[scale=0.5]{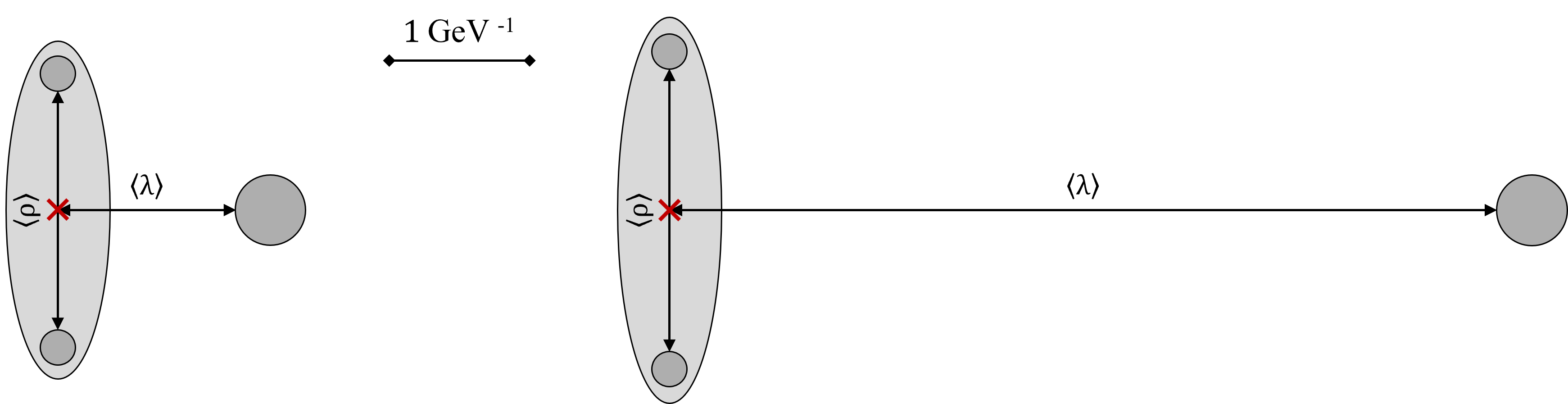}
\caption{Structures of the ground (left) and $L=8$ (right) states of the $nnb$ baryon obtained by computing the mean values of $\rho$ and $\lambda$. Both chosen states have a total spin $S=1/2$, and a total isospin $I=0$. The ratio defined in \eqref{eq: eta} takes the values $\eta=1.344$ and $\eta=0.342$ for these states, respectively.}
\label{fig: nnb fund&L=8}
\end{figure}

Although we identified a limitation of the quark-diquark approximation, the characteristic distances obtained with the three-body model remain useful to visualise the baryon structure. They allow us to study how the internal structure of the baryon affects the mass accuracy of the quark-diquark approximation. More precisely, it would be intuitive to expect that this approach provides more accurate results when the diquark is compact and the third quark orbits far from it. However, as we will see, our results indicate that the diquark compactness does not have a significant impact on the mass accuracy. To explore this further, some of the most specific baryon structures are first illustrated. Note that these representations do not take into account the full complexity of quantum systems. They are only employed as visualisation tools. In the same spirit, a large circle denotes a $b$ quark, while a small circle denotes an $n$ quark. A very detailed study of the internal structure of baryons is given in \cite{flec88}. Figures \ref{fig: bbb fund&L=8} to \ref{fig: nnb fund&L=8} are obtained with the three-body model.

Figure \ref{fig: bbb fund&L=8} displays the internal structures of the ground and $L=8$ states of the $bbb$ baryon. The configurations of these systems seem symmetric. To check this intuition, we compute the ratio
\begin{equation}
\eta=\frac{\langle \rho \rangle}{\langle \lambda \rangle},
\label{eq: eta}
\end{equation}
which equals $2/\sqrt{3} \approx 1.155$ for an equilateral triangle. We obtain $\eta=1.152$ for the ground state and $\eta=1.105$ for the $L=8$ state, confirming an approximate symmetry. This is consistent with the fact that $bbb$ baryons consist of three identical quarks. A similar symmetry is observed for $nnn$ baryons (see Fig. \ref{fig: nnn fund}). These systems also provide a useful framework to study the influence of spin interactions on baryon structures. Fig. \ref{fig: nnn fund} shows two spin states of $nnn$ baryons with clearly distinct features: the left panel represents a state with $S=3/2$, characterised by repulsive spin interactions, while the right panel shows a state with $S=1/2$, where spin interactions are either attractive or repulsive. It is clear that repulsive spin interactions tend to increase the separation between quarks, whereas attractive ones reduce it. The spin interactions are inversely proportional to the quark masses, their impact is therefore greater for light quarks. We checked that the impact of the spin-colour interactions on $bbb$ baryons is negligible.

Regarding $bbn$ and $nnb$ baryons, the right panels of Figs. \ref{fig: bbn fund&L=8} and \ref{fig: nnb fund&L=8} respectively show an extended and a compact diquark for $L=8$ states. The impact of these structures on the mass accuracy is discussed in the next section. These particular structures can be explained by energy considerations. A lower mass is obtained for the baryon by putting the high angular momentum on a two-body system containing the most massive particles in order to reduce the contribution of the centrifugal energy. Such a phenomenon has already been observed in \cite{flec88}. When two light quarks are present, the colour-spin interaction can favour a compact light diquark if it is attractive. The left panels of these figures display ground states of $bbn$ and $nnb$ baryons for comparison.

\subsection{Baryon masses} \label{subsec: Baryon masses}
The comparison between the three-body and quark-diquark approaches with $V_{Dq2}$ allows us to study the latter's accuracy in various situations of mass computation. Tables \ref{tab: M bbb} to \ref{tab: M nnb} reuse the results of Table \ref{tab: Comparison conv}. They contain the quark-diquark and three-body masses of the ground states and $L=8$ excited states of the baryons $bbb$, $nnn$, $bbn$ and $nnb$, respectively. All the relative differences between the two models seem perfectly acceptable for baryon mass predictions.

Let us first look at Table \ref{tab: M bbb} for $bbb$ baryons. In the three-body approach, the $L=0$, $S=3/2$ state has a clearly dominant component. So, its quantum numbers are used to define the state of the equivalent diquark. A very good agreement is then found. For the $L=8$, $S=1/2$ state, two components are dominant because of the symmetry constraint. One set of quantum numbers gives a better result, but differences are small. No compact diquark is present, as can be seen on Fig. \ref{fig: bbb fund&L=8}.

{\renewcommand{\arraystretch}{2}
{\setlength{\tabcolsep}{5mm}
\begin{table}[!ht]
\centering
\begin{tabular}{|c|c|c|c|c|}
\hline
$L$         & $S$                   & Approach               & Dominant state                                                                       & Mass \\ \hline \hline
\multirow{2}{*}{$0$} & \multirow{2}{*}{$\frac{3}{2}$} & 3-body                   & $-0.92(0,0,0,0,1,0)$                                                                         & $14.410$                                                                                         \\ \cline{3-5} 
                     &                                & quark-diquark                  & $(0,0,0,0,1,0)$                                                                              & $14.570\;(1.1)$                                                                                    \\ \hline
\multirow{3}{*}{$8$} & \multirow{3}{*}{$\frac{1}{2}$} & 3-body                   & \begin{tabular}[c]{@{}c@{}}$0.86\;[0.57(0,0,0,8,1,0)$\\ $-0.54(0,7,0,1,0,0)+...]$\end{tabular} & $16.488$                                                                                         \\ \cline{3-5} 
                     &                                & \multirow{2}{*}{quark-diquark} & $(0,0,0,8,1,0)$                                                                              & $16.468\;(0.1)$                                                                                   \\ \cline{4-5} 
                     &                                &                                & $(0,7,0,1,0,0)$                                                                              & $16.768\;(1.7)$                                                                                   \\ \hline
\end{tabular}
\caption{Comparison of the masses (in GeV) of different states of the $bbb$ baryon obtained using the three-body and the quark-diquark approaches. The total isospin is $I=0$. The relative differences $(\%)$ between the two models are indicated in parentheses. The first number in the 4th column is the coefficient of the state, or the coefficient of the combination of states, computed with the three-body approach. The numbers in the square brackets are the computed coefficients ensuring a good global symmetry in the three-body approach.}
\label{tab: M bbb}
\end{table}
}}

Similar situations appear for $nnn$ baryons, as can be seen on Table \ref{tab: M nnn}. But relative differences are greater than for $bbb$ baryons. This is due to the fact that relative errors are computed with the baryon mass $M_B$ and not with the binding energy $E_B = M_B-3 m_q$. Disagreements are similar for $E_B$ in $bbb$ and $nnn$ baryons, but the presence of $3 m_b$ in the mass $M_B$ reduces the relative errors. We nevertheless prefer to look at the masses because this quantity is observable and because the quark masses are also parameters of the model. A source of error can also be due to a too simple treatment of the spin-colour interactions. Indeed, the $I=S=1/2$ state with a mix of attractive and repulsive spin-colour interactions presents the largest errors. This will be discussed in the conclusion. Reasonable masses are found in the quark-diquark approach for $L=0$ states, although these systems do not contain a compact diquark, as can be seen in Fig. \ref{fig: nnn fund}.

{\renewcommand{\arraystretch}{2}
{\setlength{\tabcolsep}{4mm}
\begin{table}[!ht]
\centering
\begin{tabular}{|c|c|c|c|c|c|}
\hline
$L$         & $S$                   & $I$                   & Approach               & Dominant State                                                                                                  & Mass \\ \hline \hline
\multirow{2}{*}{$0$} & \multirow{2}{*}{$\frac{3}{2}$} & \multirow{2}{*}{$\frac{3}{2}$} & 3-body & $0.94(0,0,0,0,1,1)$                                                                                                     & $1.226$                                                    \\ \cline{4-6} 
                     &                                &                                & quark-diquark                  & $(0,0,0,0,1,1)$                                                                                                         & $1.302\; (6.2)$                                            \\ \hline
\multirow{3}{*}{$0$} & \multirow{3}{*}{$\frac{1}{2}$} & \multirow{3}{*}{$\frac{1}{2}$} & 3-body                   & \begin{tabular}[c]{@{}c@{}}$0.89\; [\frac{1}{\sqrt{2}}(0,0,0,0,0,0)$\\ $+\frac{1}{\sqrt{2}}(0,0,0,0,1,1)]$\end{tabular} & $0.954$                                                    \\ \cline{4-6} 
                     &                                &                                & \multirow{2}{*}{quark-diquark} & $(0,0,0,0,0,0)$                                                                                                         & $1.100\; (15.3)$                                           \\ \cline{5-6} 
                     &                                &                                &                                & $(0,0,0,0,1,1)$                                                                                                         & $1.192\; (24.9)$                                           \\ \hline
\multirow{3}{*}{$8$} & \multirow{3}{*}{$\frac{1}{2}$} & \multirow{3}{*}{$\frac{1}{2}$} & 3-body                   & \begin{tabular}[c]{@{}c@{}}$-0.91\; [-0.58(0,0,0,8,0,0)$\\ $-0.48(0,6,0,2,1,1)+...]$\end{tabular}                       & $3.479$                                                    \\ \cline{4-6} 
                     &                                &                                & \multirow{2}{*}{quark-diquark} & $(0,0,0,8,0,0)$                                                                                                         & $3.642\; (4.7)$                                            \\ \cline{5-6} 
                     &                                &                                &                                & $(0,6,0,2,1,1)$                                                                                                         & $3.897\; (12.0)$                                            \\ \hline
\end{tabular}
\caption{Same as Table \ref{tab: M bbb} but for $nnn$. The total isospin $I$ is indicated in the third column.}
\label{tab: M nnn}
\end{table}
}}

Tables \ref{tab: M bbn} and \ref{tab: M nnb} present the results for $bbn$ and $nnb$ baryons, respectively. Again, relative errors between the three-body and the quark-diquark approaches are small. But, as no global symmetry on the three quarks is required for these systems, new situations appear. The diquark compactness impact on the baryon masses can then be studied from a different perspective. To this end, we consider the $L=8$ states of $bbn$ and $nnb$, seeing that the former presents an extended diquark and the latter a compact one. We could expect a better correspondence between the quark-diquark and three-body masses for $nnb$, but the results clearly show that it is not the case. Indeed, the relative difference in masses for the extended diquark is lower ($1.0\%$) than for the compact diquark ($3.3\%$). These observations indicate that the diquark compactness does not guarantee a better accuracy for the quark-diquark approximation.

Before concluding, let us briefly compare the two identification methodologies mentioned at the beginning of the present section. Both methodologies are aligned for all the states: the dominant configurations $(n_D, l_D, n_{Dq}, l_{Dq}, S_D, I_D)$ chosen in Tables~\ref{tab: M bbb}, \ref{tab: M nnn}, \ref{tab: M bbn} and \ref{tab: M nnb} and selected using the procedure based on the internal structure are those that yield the ground state in the quark-diquark picture. This agreement is less obvious for the $L=8$ states of $bbb$ and $nnn$ because their wave functions contain many non-negligible configurations. However, the first dominant configuration of these states always corresponds to the lowest quark-diquark mass. Overall, both methodologies result in comparable accuracies for the baryon masses. Therefore, the quark-diquark approximation can be used to estimate masses without requiring the knowledge of the full three-body solution by resorting to energy ordering. However, the current study preferred an identification based on the internal structure to make the comparison of observables as relevant as possible.

{\renewcommand{\arraystretch}{2}
{\setlength{\tabcolsep}{5.5mm}
\begin{table}[!ht]
\centering
\begin{tabular}{|c|c|c|c|c|}
\hline
$L$         & $S$                   & Approach & Dominant State & Mass \\ \hline \hline
\multirow{2}{*}{$0$} & \multirow{2}{*}{$\frac{3}{2}$} & 3-body     & $0.92(0,0,0,0,1,0)$    & $10.224$                                                   \\ \cline{3-5} 
                     &                                & quark-diquark    & $(0,0,0,0,1,0)$        & $10.249\; (0.2)$                                           \\ \hline
\multirow{2}{*}{$0$} & \multirow{2}{*}{$\frac{1}{2}$} & 3-body     & $0.92(0,0,0,0,1,0)$    & $10.188$                                                   \\ \cline{3-5} 
                     &                                & quark-diquark    & $(0,0,0,0,1,0)$        & $10.217\; (0.3)$                                           \\ \hline
\multirow{2}{*}{$8$} & \multirow{2}{*}{$\frac{1}{2}$} & 3-body     & $-0.97(0,8,0,0,1,0)$   & $11.922$                                                   \\ \cline{3-5} 
                     &                                & quark-diquark    & $(0,8,0,0,1,0)$        & $12.041\; (1.0)$                                           \\ \hline
\end{tabular}
\caption{Same as Table \ref{tab: M bbb} but for $bbn$ and $I=1/2$.}
\label{tab: M bbn}
\end{table}
}}

{\renewcommand{\arraystretch}{2}
{\setlength{\tabcolsep}{5mm}
\begin{table}[!ht]
\centering
\begin{tabular}{|c|c|c|c|c|c|}
\hline
$L$         & $S$                   & $I$         & Approach & Dominant State & Mass \\ \hline \hline
\multirow{2}{*}{$0$} & \multirow{2}{*}{$\frac{3}{2}$} & \multirow{2}{*}{$1$} & 3-body     & $0.94(0,0,0,0,1,1)$    & $5.837$                                                    \\ \cline{4-6} 
                     &                                &                      & quark-diquark    & $(0,0,0,0,1,1)$        & $6.036\; (3.4)$                                            \\ \hline
\multirow{2}{*}{$0$} & \multirow{2}{*}{$\frac{1}{2}$} & \multirow{2}{*}{$1$} & 3-body     & $-0.94(0,0,0,0,1,1)$   & $5.812$                                                    \\ \cline{4-6} 
                     &                                &                      & quark-diquark    & $(0,0,0,0,1,1)$        & $6.027\; (3.7)$                                            \\ \hline
\multirow{2}{*}{$0$} & \multirow{2}{*}{$\frac{1}{2}$} & \multirow{2}{*}{$0$} & 3-body     & $-0.93(0,0,0,0,0,0)$   & $5.583$                                                    \\ \cline{4-6} 
                     &                                &                      & quark-diquark    & $(0,0,0,0,0,0)$        & $5.850\; (4.8)$                                            \\ \hline
\multirow{2}{*}{$8$} & \multirow{2}{*}{$\frac{1}{2}$} & \multirow{2}{*}{$0$} & 3-body     & $-0.97(0,0,0,8,0,0)$   & $7.638$                                                    \\ \cline{4-6} 
                     &                                &                      & quark-diquark    & $(0,0,0,8,0,0)$        & $7.892\; (3.3)$                                            \\ \hline
\end{tabular}
\caption{Same as Table \ref{tab: M nnn} but for $nnb$.}
\label{tab: M nnb}
\end{table}
}}

\section{Conclusion} \label{sec: Conclusion}

In this paper, we have examined the validity of the quark-diquark approximation within the framework of a QCD-inspired potential quark model. For this purpose, masses of various baryons have been computed with a semi-relativistic three-body Hamiltonian and compared with the masses obtained in a quark-diquark approximation scheme of the same Hamiltonian. We focused on baryons composed of, at least, two $n$ ($u$ or $d$) and $b$ quarks to study the influence of masses. Ground states and $L=8$ states are studied to assess the influence of the angular momentum. Other possible quark contents or intermediate angular momenta have not been considered to lighten the paper. We think that the extreme cases studied here are sufficient to draw relevant conclusions. The association of a three-body state with a quark-diquark one relies on information coming from the internal structure of the three-body state, instead of the simple order number in both spectra. We think this procedure allows a more relevant comparison.

The main result is that better baryon masses can be computed within the quark-diquark approximation using parameters coming from a three-body model if the size of the diquark is taken into account with a physically appropriate density of quarks. Some possible improvements are discussed below. We do not conclude that quark-diquark calculations are irrelevant if the size of the diquark is not taken into account or if it is taken into account in a different way than ours. It is possible to obtain good results if the parameters of the quark-diquark approximation used are properly fitted. This has been shown in numerous studies. 

If the quark-diquark approach can provide mass approximations up to a few percent for almost all the studied baryons, the quality of the internal structure is more questionable, taking into account the results obtained for inter-quark distances. This does not mean that it cannot be used to compute some observables. It is certainly necessary to adapt the definition of observables to the peculiarities of the quark-diquark structure, in particular the size of the diquark. But this is beyond the scope of this work.

We also showed that a quark-diquark description is possible even if the diquark is not a compact object. If the diquark quantum numbers are properly identified for a given baryon and its density correctly taken into account, the mass of the three-body system can be well reproduced, even if the size of the diquark is greater than the mean distance between the third quark and the diquark. These results are in agreement with the conclusions of \cite{zhu25}, showing that the separation between the two quarks in the diquark cannot be neglected. In this last paper, it is also shown that the use of the meson exchanges of the chiral quark model replacing the one gluon exchanges used in our paper just slightly increases the diquark size in most systems. 

The choice of the quark-diquark configuration to compute the mass of a given baryon can be guided by a minimal energy condition. This also holds for excited states whose three-body wave functions involve many quark-diquark configurations with significant probabilities, where at least one of the dominant configurations corresponds to the lowest quark-diquark mass. In general, the minimal energy condition does not substantially deteriorate the mass accuracy compared to our method.

If all the richness and the complexity of a three-body system cannot be fully reproduced by a model with a preformed two-body sized cluster, we think that our results may be useful for studies of more complex systems, such as tetraquarks \cite{carl08}, pentaquarks \cite{gian19}, hexaquarks \cite{an25}, and even hybrid baryons \cite{cimi24}, where substructures like diquarks or triquarks are considered. It is also possible to build models in which the quark-diquark approach is merged with a usual three-quark model \cite{gala12}. Nevertheless, this work highlights some limitations of the quark-diquark approximation, and shows that it can be improved. We have identified three points of concern.

In a quark-diquark system consisting of three identical quarks, several spatial configurations are compatible with the Pauli principle, but only one can be selected for the total wave function. This problem can be solved by allowing a mixing of configurations for the diquark instead of a unique frozen state. This is possible since the mixing coefficient can be computed with the group theory.

It is also the case when a $qqQ$ system is studied with a $(qQ)$-$q$ substructure instead of $(qq)$-$Q$, as it is always done in this work. This last problem seems relevant to study, since we have some indications that a $(nb)$-$n$ configuration could be energetically more favourable than a $(nn)$-$b$ one. To examine this problem in detail, it is first necessary to generalise our quark density to the case of two different particles. Calculations are feasible but will be more complicated because, in this case, the position of the centre of mass of two different quarks depends on the momentum of the quarks.

A third possible improvement is to better take into account the spin-colour contribution. Our way of computing this interaction between the quark and the diquark is equivalent to considering that the two spins inside the diquark are simply added. A better way may be to compute the colour-magnetic moment of the diquark, taking into account its internal structure. Several possibilities can be investigated \cite{maje16,lich83}.

\section*{Acknowledgements}
C.T. has received support from the Communauté française de Belgique as part of the funding for a FRIA grant. C.C. would like to thank the Fonds de la Recherche Scientifique - FNRS for the financial support. This work was also supported by the IISN under Grant Number 4.45.10.08.

\clearpage
\bibliography{bibliography}

\clearpage
\appendix
\section{Characteristic distances results} \label{app: Characteristic distances results}

{\renewcommand{\arraystretch}{2}
{\setlength{\tabcolsep}{5mm}
\begin{table}[!ht]
\centering
\begin{tabular}{|c|c|c|c|c|c|}
\hline
$L$         & $S$                   & Approach               & Dominant state                                                                         & $\braket{\rho}$ or $r_{qq}$ & $\braket{\lambda}$ or $r_{Dq}$ \\ \hline \hline
\multirow{2}{*}{$0$} & \multirow{2}{*}{$\frac{3}{2}$} & 3-body                  & $-0.92(0,0,0,0,1,0)$                                                                           & $0.742$           & $0.644$           \\ \cline{3-6} 
                     &                                & quark-diquark                  & $(0,0,0,0,1,0)$                                                                                & $1.052\; (41.8)$  & $0.767\; (19.1)$  \\ \hline
\multirow{3}{*}{$8$} & \multirow{3}{*}{$\frac{1}{2}$} & 3-body                  & \begin{tabular}[c]{@{}c@{}}$0.86\;[0.57(0,0,0,8,1,0)$\\ $-0.54(0,7,0,1,0,0)+...]$\end{tabular} & $3.260$           & $2.951$           \\ \cline{3-6} 
                     &                                & \multirow{2}{*}{quark-diquark} & $(0,0,0,8,1,0)$                                                                                & $1.052\; (67.7)$  & $4.350\; (47.4)$  \\ \cline{4-6} 
                     &                                &                                & $(0,7,0,1,0,0)$                                                                                & $5.716\; (75.3)$  & $2.122\; (28.1)$  \\ \hline
\end{tabular}
\caption{Comparison of the characteristic distances (in GeV$^{-1}$) of different states of the $bbb$ baryon obtained using the three-body, $\braket{\rho}$ and $\braket{\lambda}$, and the quark-diquark approaches, $r_{qq}$ and $r_{Dq}$. The total isospin is $I=0$. The relative differences $(\%)$ between the two models are indicated in parentheses. The first number in the 4th column is the coefficient of the state, or the coefficient of the combination of states, computed with the three-body approach. The numbers in the square brackets are the computed coefficients ensuring a good global symmetry in the three-body approach.}
\label{tab: r bbb}
\end{table}
}}

{\renewcommand{\arraystretch}{2}
{\setlength{\tabcolsep}{4mm}
\begin{table}[!ht]
\centering
\begin{tabular}{|c|c|c|c|c|c|c|}
\hline
$L$         & $S$                   & $I$                   & Approach               & Dominant state                                                                                                  & $\braket{\rho}$ or $r_{qq}$ & $\braket{\lambda}$ or $r_{Dq}$ \\ \hline \hline
\multirow{2}{*}{$0$} & \multirow{2}{*}{$\frac{3}{2}$} & \multirow{2}{*}{$\frac{3}{2}$} & 3-body                    & $0.94(0,0,0,0,1,1)$                                                                                                     & $2.685$           & $2.329$           \\ \cline{4-7} 
                     &                                &                                & quark-diquark                  & $(0,0,0,0,1,1)$                                                                                                         & $3.495\; (30.2)$  & $2.529\; (8.6)$   \\ \hline
\multirow{3}{*}{$0$} & \multirow{3}{*}{$\frac{1}{2}$} & \multirow{3}{*}{$\frac{1}{2}$} & 3-body                    & \begin{tabular}[c]{@{}c@{}}$0.89\; [\frac{1}{\sqrt{2}}(0,0,0,0,0,0)$\\ $+\frac{1}{\sqrt{2}}(0,0,0,0,1,1)]$\end{tabular} & $2.186$           & $1.900$           \\ \cline{4-7} 
                     &                                &                                & \multirow{2}{*}{quark-diquark} & $(0,0,0,0,0,0)$                                                                                                         & $2.800\; (28.1)$  & $2.386\; (25.6)$  \\ \cline{5-7} 
                     &                                &                                &                                & $(0,0,0,0,1,1)$                                                                                                         & $3.495\; (59.9)$  & $2.333\; (22.8)$  \\ \hline
\multirow{3}{*}{$8$} & \multirow{3}{*}{$\frac{1}{2}$} & \multirow{3}{*}{$\frac{1}{2}$} & 3-body                    & \begin{tabular}[c]{@{}c@{}}$-0.91\; [-0.58(0,0,0,8,0,0)$\\ $-0.48(0,6,0,2,1,1)]$\end{tabular}                           & $5.915$           & $5.297$           \\ \cline{4-7} 
                     &                                &                                & \multirow{2}{*}{quark-diquark} & $(0,0,0,8,0,0)$                                                                                                         & $2.800\; (52.7)$  & $8.097\; (52.9)$  \\ \cline{5-7} 
                     &                                &                                &                                & $(0,6,0,2,1,1)$                                                                                                         & $10.210\; (72.6)$ & $4.755\; (10.2)$  \\ \hline
\end{tabular}
\caption{Same as Table \ref{tab: r bbb} but for $nnn$. The total isospin $I$ is indicated in the third column.}
\label{tab: r nnn}
\end{table}
}}

{\renewcommand{\arraystretch}{2}
{\setlength{\tabcolsep}{5.5mm}
\begin{table}[!ht]
\centering
\begin{tabular}{|c|c|c|c|c|c|}
\hline
$L$         & $S$                   & Approach & Dominant state & $\braket{\rho}$ or $r_{qq}$ & $\braket{\lambda}$ or $r_{Dq}$ \\ \hline \hline
\multirow{2}{*}{$0$} & \multirow{2}{*}{$\frac{3}{2}$} & 3-body      & $0.92(0,0,0,0,1,0)$    & $0.917$           & $1.407$           \\ \cline{3-6} 
                     &                                & quark-diquark    & $(0,0,0,0,1,0)$        & $1.052\; (14.7)$  & $1.484\; (5.5)$  \\ \hline
\multirow{2}{*}{$0$} & \multirow{2}{*}{$\frac{1}{2}$} & 3-body      & $0.92(0,0,0,0,1,0)$    & $0.905$           & $1.352$           \\ \cline{3-6} 
                     &                                & quark-diquark    & $(0,0,0,0,1,0)$        & $1.052\; (16.2)$  & $1.437\; (6.3)$  \\ \hline
\multirow{2}{*}{$8$} & \multirow{2}{*}{$\frac{1}{2}$} & 3-body      & $-0.97(0,8,0,0,1,0)$   & $4.901$           & $2.384$           \\ \cline{3-6} 
                     &                                & quark-diquark    & $(0,8,0,0,1,0)$        & $6.187\; (26.2)$  & $2.523\; (5.8)$   \\ \hline
\end{tabular}
\caption{Same as Table \ref{tab: r bbb} but for $bbn$ and $I=1/2$.}
\label{tab: r bbn}
\end{table}
}}

{\renewcommand{\arraystretch}{2}
{\setlength{\tabcolsep}{5mm}
\begin{table}[!ht]
\centering
\begin{tabular}{|c|c|c|c|c|c|c|}
\hline
$L$         & $S$                   & $I$         & Approach & Dominant state & $\braket{\rho}$ or $r_{qq}$ & $\braket{\lambda}$ or $r_{Dq}$ \\ \hline \hline
\multirow{2}{*}{$0$} & \multirow{2}{*}{$\frac{3}{2}$} & \multirow{2}{*}{$1$} & 3-body      & $0.94(0,0,0,0,1,1)$    & $2.286$           & $1.480$           \\ \cline{4-7} 
                     &                                &                      & quark-diquark    & $(0,0,0,0,1,1)$        & $3.495\; (52.9)$  & $1.851\; (25.1)$  \\ \hline
\multirow{2}{*}{$0$} & \multirow{2}{*}{$\frac{1}{2}$} & \multirow{2}{*}{$1$} & 3-body      & $-0.94(0,0,0,0,1,1)$   & $2.250$           & $1.446$           \\ \cline{4-7} 
                     &                                &                      & quark-diquark    & $(0,0,0,0,1,1)$        & $3.495\; (55.3)$  & $1.839\; (27.2)$  \\ \hline
\multirow{2}{*}{$0$} & \multirow{2}{*}{$\frac{1}{2}$} & \multirow{2}{*}{$0$} & 3-body      & $-0.93(0,0,0,0,0,0)$   & $1.706$           & $1.269$           \\ \cline{4-7} 
                     &                                &                      & quark-diquark    & $(0,0,0,0,0,0)$        & $2.800\; (64.1)$  & $1.762\; (38.8)$  \\ \hline
\multirow{2}{*}{$8$} & \multirow{2}{*}{$\frac{1}{2}$} & \multirow{2}{*}{$0$} & 3-body      & $-0.97(0,0,0,8,0,0)$   & $2.007$           & $5.877$           \\ \cline{4-7} 
                     &                                &                      & quark-diquark    & $(0,0,0,8,0,0)$        & $2.800\; (39.5)$  & $6.416\; (9.2)$   \\ \hline
\end{tabular}
\caption{Same as Table \ref{tab: r nnn} but for $nnb$.}
\label{tab: r nnb}
\end{table}
}}

\end{document}